\documentclass[pra,twocolumn]{revtex4-1}

\usepackage{subfigure}
\usepackage{dcolumn}

\usepackage{graphicx}

\newcommand{\cascade}{\textsc{cascade}}

\begin{document}

\title{High Performance Information Reconciliation for QKD with CASCADE}
\author{Thomas Brochmann \surname{Pedersen}}
\email{thomas.pedersen@tubitak.gov.tr}
\author{Mustafa Toyran} 
\email{mustafa.toyran@tubitak.gov.tr}
\affiliation{Quantum Cryptology Division,  B\.ILGEM, T\"UB\.ITAK, Turkey}
\date{\today}

\begin{abstract}
  It is widely accepted in the quantum cryptography community that
  interactive information reconciliation protocols, such as
  \cascade{}, are inefficient due to the communication
  overhead. Instead, non-interactive information reconciliation
  protocols based on i.e. LDPC codes or, more recently, polar codes
  have been proposed. In this work, we argue that interactive
  protocols should be taken into consideration in modern quantum key
  distribution systems. In particular, we demonstrate how to improve
  the performance of \cascade{} by proper implementation and use. Our
  implementation of \cascade{} reaches a throughput above 80~Mbps
  under realistic conditions. This is more than four times the
  throughput previously demonstrated in any information reconciliation
  protocol.
\end{abstract}

\maketitle

\section{Introduction}

Information reconciliation (IR) in quantum key distribution (QKD) is a
protocol where Alice and Bob, by public discussion over an
authenticated classical channel, correct the discrepancies between the
bit strings which they obtained through usage of the noisy quantum
channel.

IR has been the bottleneck in many QKD systems, both
discrete\cite{Toshiba08,ninoqcrypt,MEM13} and
continuous\cite{CNRS07,CNRS11b} variable. In this paper, we address
the IR problem for the binary symmetric channel, which is the model
used for discrete variable QKD.

There are two main measures of performance of an IR protocol:
Efficiency (e.g. the ability to correct the discrepancies without
revealing more information than necessary to an eavesdropper) and
throughput (e.g. how many input bits per second can be processed).

The \cascade{} IR protocol\cite{BS93} is simple and probably the most
widely used in QKD implementations. In \cascade{}, Alice and Bob first
permute and partition a frame of bits. They then compare the parities
of each partition. When they disagree on the parity of a partition,
the partition is split into two and the parities of the two halves are
compared. The half where the parities disagree is then recursively
split and checked until the error is found and corrected. This
procedure continues for a few rounds with different permutations and
partition sizes. As is easily seen, \cascade{} is highly interactive
which makes it very sensitive to network latencies. It is commonly
believed that the interaction in \cascade{} causes low
throughput\cite{CNRS13,NIST12,Winnow,MEM13}. In contrast to
\cascade{}, modern IR protocols based on forward error correction
methods, such as LDPC or polar codes, are non-interactive. They do,
however, require more computation than \cascade{}.

Efficiency is the performance measure most commonly addressed in works
on IR (See
i.e. \cite{MEM13b,CNRS08,Yan08,SY00,BS93,Winnow,LTD03}). However, as
argued in \cite{MEM13,CNRS13}, the trade-off between communication and
computation costs must be carefully evaluated when choosing IR
protocols. As long as the IR has a higher throughput than the provided
raw-key rate, efficiency is the dominating performance criteria.
However, if the IR throughput is low, a combination of the throughput
and efficiency determines the performance of the whole QKD system.
 
In low latency networks, it is not a priori clear whether
computationally simple but interactive IR protocols will perform worse
than computationally complex but non-interactive protocols. The aim of
this work is to challenge the assumption that interactive IR
protocols, such as \cite{BS93,Winnow,LTD03}, have worse performance
than non-interactive ones. To support our claim, we demonstrate that
\cascade{}, by proper implementation and usage, can outperform state
of the art IR protocols\cite{CNRS13,MEM13,ninoqcrypt,NIST07,NIST12}
for many settings which are relevant to QKD, such as QKD over fiber
channels.

In most realistic deployments of QKD, the authenticated classical
channel will have a latency of at most a couple of milliseconds. In a
QKD system over fiber, it is fair to assume that the classical channel
is either multiplexed with the quantum
channel\cite{Toshiba12,ninoqcrypt} or sent over another fiber in the
same fiber bundle. For the distances typical for a QKD system, a
direct fiber connection will give latencies close to $1$~ms. Even for
free-space QKD with low earth orbit satellites\cite{Fuchs+13}, the
latency will not exceed more than a couple of milliseconds. We
demonstrate how \cascade{} can obtain a throughput above $80$~Mbps in
the very common low latency scenarios. Only when the network latency
exceeds 10~milliseconds, does the throughput of our implementation of
\cascade{} become too low for state-of-the-art QKD systems. A
potential setting where the latency is high enough to rule out
\cascade{} is QKD with a geostationary satellite where the latency
will be a few hundred milliseconds.

This paper is organized as follows: We give the theoretical background
of IR performance in Section~\ref{sec:key-rate}. To justify the
parameters which we use in our tests, Section~\ref{sec:qkd} lists a
few state-of-the-art QKD systems and their requirements for the IR
protocol. We give a brief outline of some of the best performing IR
protocols in Section~\ref{sec:ir}, and a description of our
implementation of the \cascade{} IR protocol in
Section~\ref{sec:cascade}. The results of our experiments are listed
in Section~\ref{sec:experiments}, followed by a few concluding remarks
in Section~\ref{sec:conclusion}.

\section{\label{sec:key-rate}Secret Key Rate}

After performing sifting and error estimation, Alice and Bob each have
a \emph{frame} of a predetermined number of bits. Alice's and Bob's
frames are represented by random variables $A$ and $B$,
respectively. During information reconciliation, Alice and Bob
exchange information which will allow Bob to compute the value of
$A$. The process should leak the smallest possible amount of
information about $A$ to the eavesdropper. It follows from the
noiseless coding theorem that the minimum amount of information which
Alice and Bob need to exchange is $H(A|B)$, leaving at most
\begin{equation}
H(A) - H(A|B) = I(A{:}B)
\end{equation}
bits of information which is unknown to an eavesdropper listening to
the communication on the classical channel.

In a practical implementation of IR, however, Alice and Bob will
exchange more than $H(A|B)$ bits of information. The \emph{efficiency}
of an IR protocol is a number $\alpha \in [0,1]$ such that Alice and
Bob can extract $\alpha I(A{:}B)$ bits when IR succeeds. A further
limitation of practical implementations of IR protocols is the
probability that the protocol fails for a given frame. This
probability is called the \emph{frame error rate} (FER).

Besides the information leaked during IR, Alice and Bob also need to
take the information which the eavesdropper obtained during their use
of the quantum channel into account. Putting together all these
factors, the maximum number of secret bits which Alice and Bob can
extract from a frame is
\begin{equation}
(1 - FER)(\alpha I(A{:}B) - I_E),
\end{equation}
where $I_E$ is a measure of the information which the eavesdropper has
obtained during the quantum part of the protocol. The value of $I_E$
depends on both the specific QKD protocol and the security proof used.

If raw key is provided at a rate of $R_s$ bits per second, and the IR
is capable of correcting at that rate, the \emph{secret key rate} is
at most
\begin{equation}
R_s(1 - FER)(\alpha I(A{:}B) - I_E)
\end{equation}
bits per second, in which case the IR must reduce the FER and improve
the efficiency in order to get the best possible utility out of the
quantum channel.

If, however, the IR protocol is only capable of correcting at a rate
of $R_{\mathrm{IR}} < R_s$ bits per second, then the IR protocol
becomes the bottle\-neck in the system. Several strategies can be used
to improve on this situation. Some IR protocols, such as the ones
based on LDPC codes, can improve the throughput by allowing a higher
frame error rate.  Other IR protocols may have a throughput-efficiency
trade-off which allow them to obtain a higher throughput by
sacrificing efficiency. If we let the variable $x$ describe the
parameters which influence the trade-offs of a given IR protocol, the
maximal secret key rate, the \emph{performance}, becomes
\begin{equation}
\label{eq:performance}
R_{\mathrm{IR}}(x)(1 - FER(x))(\alpha(x) I(A{:}B) - I_E)
\end{equation}
bits per second, where $R_{\mathrm{IR}}$, $FER$, and $\alpha$ are all
functions of the parameters, $x$. In \cascade{}, for instance,
changing partition sizes will change both throughput and
efficiency. In the belief propagation decoders commonly used in LDPC,
the number of iterations gives a trade-off between throughput and
FER. In the scenario of Eq.~(\ref{eq:performance}), where IR
throughput is lower than raw-key rate, the trade-off must be carefully
evaluated to find the maximum secret key rate.

\section{\label{sec:qkd}State of the Art QKD}

The choice of information reconciliation protocol and trade-offs
between efficiency, throughput, and frame error rate depends on the
raw-key rate, the latency and bandwidth of the classical channel, and
the quantum bit error rate (QBER). The aim of this paper is to
demonstrate that interactive IR protocols (in particular \cascade{})
should be taken into consideration in current state-of-the-art QKD
systems. To support our thesis, we list the properties of some of the
state-of-the-art QKD systems.

Several recent QKD experimental setups have reached secret key rates
of up to 1~Mbps\cite{Toshiba12,WalentaThesis,NIST06}.

In a series of papers\cite{Toshiba08,Toshiba10,Toshiba12}, the group
at Toshiba Research documents progress with their 1~Mbps QKD system
based on the BB84 protocol\cite{BB84} with decoy
states\cite{H03}. While the secret key rate has been stable at 1~Mbps,
the distance has increased from 20~km\cite{Toshiba08} to
50~km\cite{Toshiba10}, and the duration of stable operation has
increased from a few seconds\cite{Toshiba08} to virtually
unlimited\cite{Toshiba10}. They use optical fiber in their work, and
experience a QBER of 2.5\%--4\% depending on the setting. The raw-key
rate is between 1.5~Mbps\cite{Toshiba12} and
3.44~Mbps\cite{Toshiba08}. An important new achievement is the
multiplexing of the quantum and classical
channels\cite{Toshiba12}. This allows cheap and very low latency
classical communication. In their older work, they used the \cascade{}
IR protocol\cite{Toshiba07}. But in their high-speed QKD
systems\cite{Toshiba08}, they identify the need for high-speed IR
protocols. In their recent work\cite{Toshiba08,Toshiba12}, the secret
key rate is simulated under the assumption of an information
reconciliation protocol with the same efficiency as \cascade{}, but
with a throughput which is fast enough to keep up with the raw-key
rate.

The group at Universit\'e de Gen\`eve has a series of QKD
implementations based on the COW\cite{SBGSZ05} protocol. Details of
the work can be found in the ph.d.\ dissertation of Nino
Walenta\cite{WalentaThesis}. At Qcrypt 2012, the group reported on a
QKD system capable of maintaining a secret key rate of $0.88$~Mbps
over $1$~km fiber\cite{ninoqcrypt}. In principle, the system should be
capable of keeping a secret key rate of $1$~Mbps for fibers of up to
$20$~km. Continuous operation of the high secret key rate was possible
due to a high-speed hardware post processing system implemented in
FPGA. The operational QBER is from 1.5\% to 2.25\%. Experiments over
$25$~km fiber showed a raw-key rate of $6.29$~Mbps over a period of 8
hours\cite{Geneva12}.

Meanwhile, the group at NIST achieved 4~Mbps raw-key rate in their
BB84 implementation over 1~km fiber in 2006\cite{NIST06}. The
classical channel used a separate 1~km fiber. The QBER was 3.42\%.

The longest QKD link reported\cite{wcg12} is $260$~km over standard
telecom fiber. The secret key rate was $1.85$~bit/s and the raw key
rate less than $100$~bit/s. Nonetheless, the QBER was still only
$3.46\%$. A classical link sent over a fiber of the same length will
have a latency of approximately $1.5$--$2$~ms (including media
converters and authentication).

As demonstrated above, state-of-the-art QKD systems require IR
protocols capable of correcting a QBER of $1.5\%$--$4\%$ with a
throughput of at least $1$--$6.29$~Mbps. The distance of fiber based
QKD systems is still limited to a few hundred kilometers at best. With
the ability to either multiplex quantum and classical channels, or
just use different fibers in the same fiber bundle, latencies will be
close to $1$~ms. The focus of our work is this range of parameters.

\section{\label{sec:ir}High Performance Information Reconciliation}

To the best of our knowledge, the highest throughput reported for an
implementation of \cascade{} is $5.5$~Mbps with a QBER of $3.8\%$ and
a frame size of $1$~Mbit\cite{TokyoNet}. The implementation uses
multiple threads on quad-core computers, and the classical channel is
sent over a 45~km optical fiber in the same fiber bundle as the
quantum channel.

In \cite{CNRS13}, the authors present IR protocols based on both LDPC
and polar codes. Their implementations are tested on a channel with
2\% QBER. They obtain the highest throughput of 10.9~Mbps with polar
codes on a 3.47GHz Intel Core i5 processor using a single core. The
block size is $2^{16}$ bits, the frame error rate is 9\%, and the
efficiency is 93.5\%. By increasing the frame size to $2^{24}$ bits,
the authors improved the efficiency to 98\%, but by sacrificing the
throughput which dropped to 8.3~Mbps. They present two implementations
of IR based on LDPC codes: One implemented on GPU (AMD Tahiti Graphics
Processor) and one on CPU (same processor as for the polar code
implementation). The GPU implementation has a throughput of 7.3~Mbps,
a FER of 1\%, and an efficiency of 92.9\%, while the CPU
implementation has a throughput of 0.83~Mbps, a FER of 3\%, and an
efficiency of 93.1\%. In both cases the block size is 131072 bits.

Another implementation of LDPC on GPU (NVIDIA GeForce GTX 670) is
presented in \cite{MEM13}. The authors achieve a throughput of up to
10.3~Mbps with a rate-adaptive LDPC code. The authors point out that
LDPC codes have a trade-off between FER, efficiency and
throughput. When the code operates close to the Shannon limit, either
a non-negligible FER or a limited throughput must be accepted.

The fastest IR protocol known to us is the LDPC code implemented on
FPGA (Xilinx Virtex 5) presented at Qcrypt 2012\cite{ninoqcrypt}. They
use standard IEEE 802.11n quasi-cyclic LDPC codes with a frame size of
1944 bits. Their IR protocol has been demonstrated to process
$20$~Mbps raw-key. They claim that it will be capable of operating at
up to $40.8$~Mbps in the future. The IR protocol has a FER of up to
$10\%$.

For their QKD system, NIST has implemented both \cascade{} and LDPC on
FPGA, achieving $5$ and $4$~Mbps, respectively\cite{NIST12}. In line
with our thesis, the authors of \cite{NIST12} indicate that \cascade{}
has a higher throughput than LDPC for distances up to at least
$100$~km. In \cite{NIST07}, Alan Mink of NIST reaches $12.2$~Mbps for
$1$\% QBER with an unspecified IR protocol by running 4 threads in
parallel on an FPGA board\footnote{Citation to \cite{NIST07} in
  \cite{NIST12} leads us to believe that the IR protocol used in
  \cite{NIST07} is \cascade{}, or a variant thereof, which would make
  it the fastest implementation of \cascade{} known to us.}.

\section{\label{sec:cascade}CASCADE}

Interaction is the main performance concern in \cascade{}. The first
natural step towards an efficient implementation of \cascade{} is
therefore the reduction of interaction. As pointed out by Louis
Salvail at a SECOQC project meeting\cite{louis,TokyoNet}, the parity checks of
partitions and sub-blocks can be done in parallel. First, the parities
of all partitions of a frame are exchanged in a single message. Then,
instead of correcting each partition with errors one by one, all
partitions with errors are split, and the parities of their sub-blocks
are sent in a single message. The binary search for errors continues
in the same fashion by splitting sub-blocks with errors and exchanging
their parities in single messages before splitting again. For a
partition size of $k$ bits, the total number of messages exchanged
during the binary search is then $\log_2(k)$ regardless of the number
of partitions and the size of the frame. This approach is also used by
the \cascade{} implementation in the AIT QKD software
project\cite{aitqkd}.

When an error is detected in round two or later, it implies that the
error was located in partitions with even numbers of errors after each
previous round. The original description of \cascade{}\cite{BS93} has
a look-back step after each round to take advantage of this fact. For
instance, consider a partition which contains two undetected errors
after the first round (Alice and Bob agreed on the parity of that
partition after round one). If one of the errors is corrected in round
two, the other error can be corrected by applying the binary search
for errors on the partition from round one (which contains exactly one
error after the first error was corrected). In look-back, for each
error corrected in a round, all partitions from previous rounds
containing that error are added to a look-back list. The binary search
for errors is then applied to each of the partitions in the look-back
list, starting with the smallest partition. For each new error
corrected during look-back, the partitions from all rounds containing
that error are also added to the look-back list. This procedure
continues until the look-back list is empty.

In the original formulation of the look-back step, only one partition
can be corrected at one time. This imposes a high level of
interaction. To overcome this problem, we propose a slight
modification to the look-back step: Instead of only applying the
binary search to the smallest partition, it is applied to all
partitions of the earliest previous round which still has partitions
which may contain errors. Since the corrections are done to
non-overlapping partitions from the same round, we can apply the
binary search in parallel as described above. During the second round,
this will exactly be the same partitions as visited during the first
look-back in the original protocol. As demonstrated in
Section~\ref{sec:experiments}, this modification does not reduce the
efficiency noticeably for a QBER below 10\%.

With our approach, the number of messages which have to be sent
between Alice and Bob is
\begin{equation}
  \label{eq:messages}
  r + \sum_{i=1}^r \left\lceil \log_2(k_i) \right\rceil + l(QBER,n,
  k_1, \ldots, k_r),
\end{equation}
where $r$ is the number of rounds, $k_i$ is the partition size used in
the $i$th round, and $l$ is a function describing the number of
messages exchanged during look-back. For the partition sizes used in
\cite{BS93}, $l$ only increases logarithmic in the frame size, $n$
(details in Section~\ref{sec:experiments}).

An important feature of equation~\ref{eq:messages} is that, again for
the standard partition sizes, it increases sub-linearly in the frame
size. To get the biggest possible advantage of this implementation,
the frame size should be as large as possible. The larger the frame
size, the more weight is moved away from the communication and onto
the computation of the protocol. After a certain frame size, most of
the time will be spend on computation (calculation of parities).

The major factor in limiting the maximum possible frame size is
memory. The memory required increases linearly with the frame size. It
is therefore important to use as little memory as possible. A first,
trivial step to save memory is to make sure that each byte of memory
is used to store 8 data bits. A more challenging situation arises
because of the look-back step. Before each round in \cascade{}, the
data frame is permuted. In look-back, we need to find the partition of
a previous round which contained the bit which has been corrected in
the current round. In particular, we need to invert the permutation of
the current round, and apply the permutation of the previous round. In
the original description of \cascade{}, permutation is done with a
hash function, $h$, from a family of universal hash functions: The bit
at index $i$ is assigned to partition number $h(i)$. However, knowing
that a bit is in partition $h(i)$ does not allow us to compute the
original index of the bit, let alone the index in a previous round,
which uses a different hash function. A simple solution is to store
the mapping from original indices to permuted indices. Storing this
mapping, however, requires at least $n\log_2(n)$ bits (at least
$232$~Mbit to store the permutation of a $10$~Mbit frame). For large
block sizes, which is our aim, the overhead is prohibitively
large. Instead, we permute the data bits with a random, invertible
function --- essentially a random number generator, where the original
index of a bit is used as seed.

Once large block sizes are used, and computation has become the
bottleneck, throughput is improved by speeding up the computation in
\cascade{}. \cascade{} does not have much computation: Mostly the
computation of parities. During the binary search for an error, after
splitting up a block of bits, the parity of one of the sub-blocks has
to be calculated: This involves revisiting half the bits which were
already used to compute the parity of the original block. To limit the
number of repeat look-ups, we compute a \emph{prefix parity} list
during the calculation of the parities of the partitions in the
beginning of a round. The $i$th element of the \emph{prefix parity}
list of the permuted data frame $[d_0, \ldots, d_{n-1}]$ is
\begin{equation}
  pp_i = pp_{i-1} \oplus d_{i-1},
\end{equation}
for $i \in \{1,\ldots,n\}$, and $pp_0 = 0$. Computing the prefix
parity list takes the same time as computing the parities of the
partitions, but once the prefix parity list is computed, the parity of
\emph{any} interval of data bits $[d_{i},\ldots,d_{i+l}]$ can be
computed by looking up only two values in the prefix parity list:
\begin{equation}
  \mathrm{parity}([d_{i},\ldots,d_{i+l}]) = pp_i \oplus pp_{i+l+1}.
\end{equation}

In the original description of the look-back step, only the smallest
partitions of previous rounds are searched for a recently corrected
bit. In \cite{Yan08}, the authors propose an improvement where
\emph{all blocks} seen during \cascade{} are used during
look-back. This decreases the expected size of the smallest block
containing the newly corrected bit, thus making the following binary
search less interactive. In our implementation of \cascade{}, we apply
this improvement.

The source code of our \cascade{} implementation is available upon
request to the principal author.

\section{\label{sec:experiments}Experimental Results}

Our implementation of \cascade{} was tested on two Intel Core i7 3.4
Ghz CPUs (comparable to the tests performed in \cite{CNRS13})
connected by a gigabit Ethernet connection. Except for tests where we
explicitly mention the number of parallel processes used, Alice and
Bob use only a single \cascade{} process on each their computer. All
test results are the average of correcting at least 100 frames.

\begin{figure}
  \includegraphics{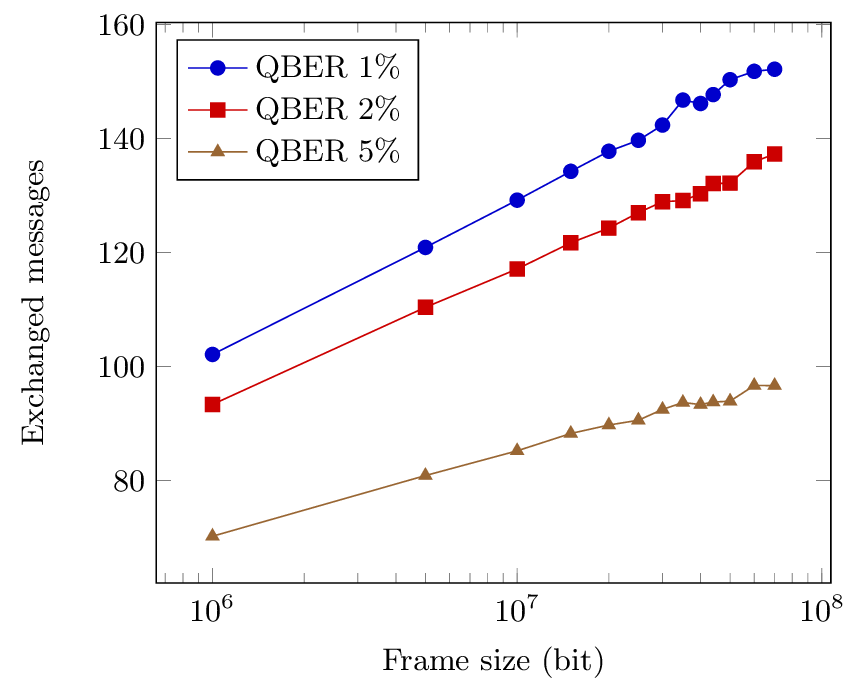}
  \caption{\label{fig:messages}The number of messages exchanged to
    correct a single frame. The dependency on frame size comes from
    the look-back procedure. The logarithmic dependency on frame size
    encourages the use of large frames.}
\end{figure}

Our first experiment addresses the ``feared'' amount of interaction of
\cascade{}. Fig.~\ref{fig:messages} shows the number of messages
exchanged between Alice and Bob as a function of frame size. As
mentioned in the previous section, the number of messages increases
logarithmic in the frame size. This sub-linearity results in an
relative drop in the cost of communication when the frame size is
increased. The figure also shows that the number of messages exchanged
decreases as QBER increases. This, in part, is due to the use of
smaller partition sizes for larger QBER, which reduces the number of
messages exchanged in each call to the binary search for errors.

To test the effect of the interaction under different network
latencies, we used facilities of the Linux kernel\cite{linuxtc} to set
the latency. All tests refer to the end-to-end latency (not round-trip
latency). The time needed to authenticate the communication is ignored
(included in the network latency) in all tests except for a single
real-world test over optical fiber described below.

\begin{figure}
  \subfigure[][QBER 1\%]{
    \includegraphics{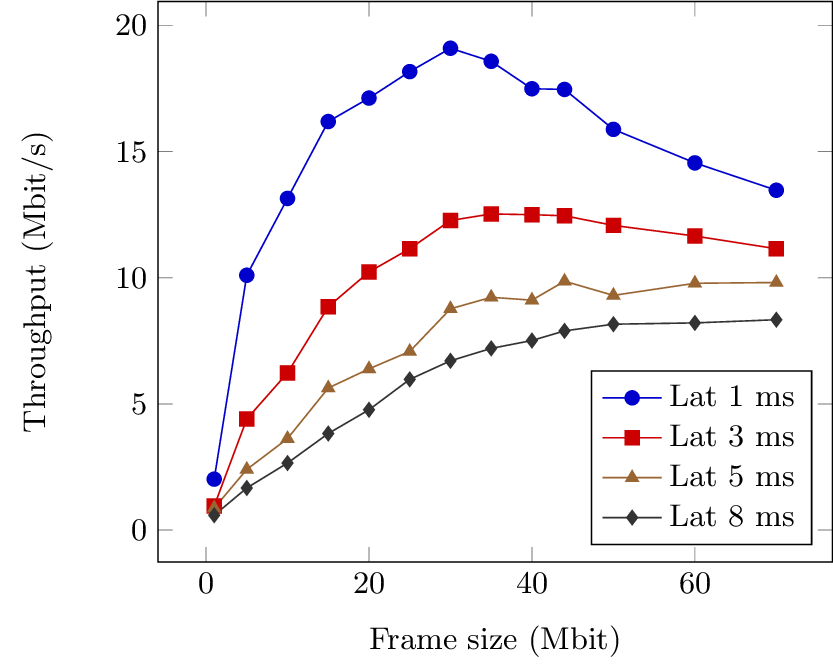}
  }  \subfigure[][QBER 5\%]{
    \includegraphics{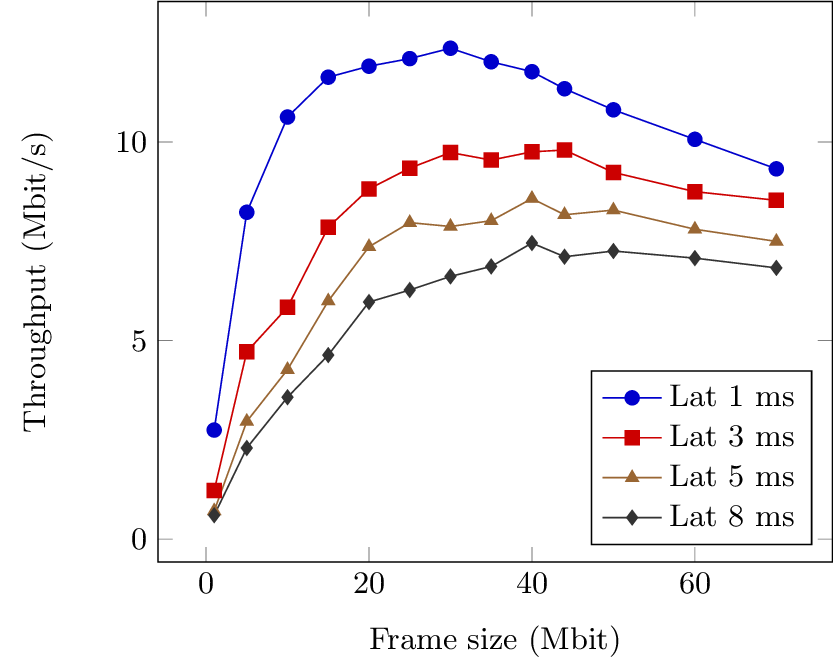}
  }
  \subfigure[][QBER 15\%]{
    \includegraphics{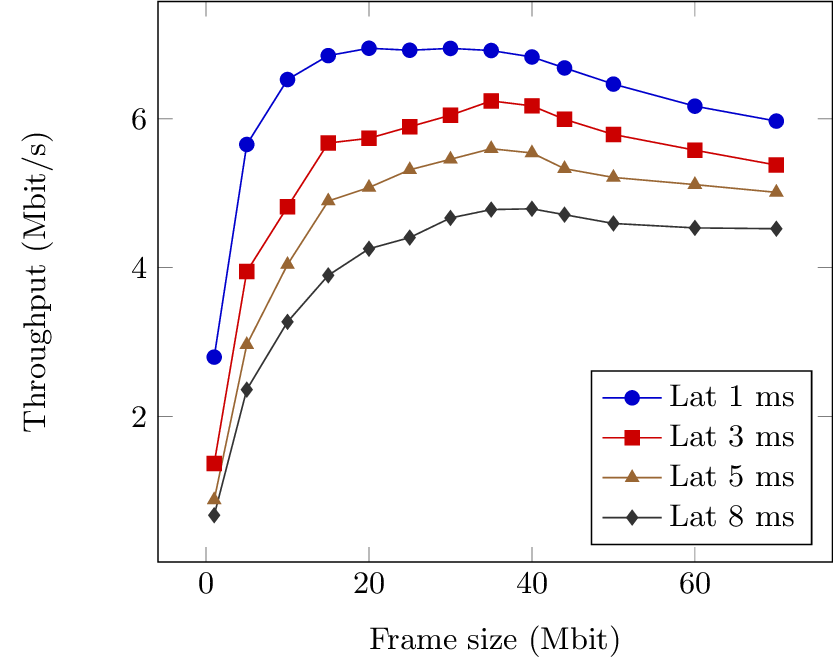}
  }
  \caption{\label{fig:tpvsn}Larger block sizes take better advantage
    of the parallel execution of the binary search for errors. After a
    certain point, the computation time overtakes the reduction of
    communication time.}
\end{figure}

\begin{figure}
    \includegraphics{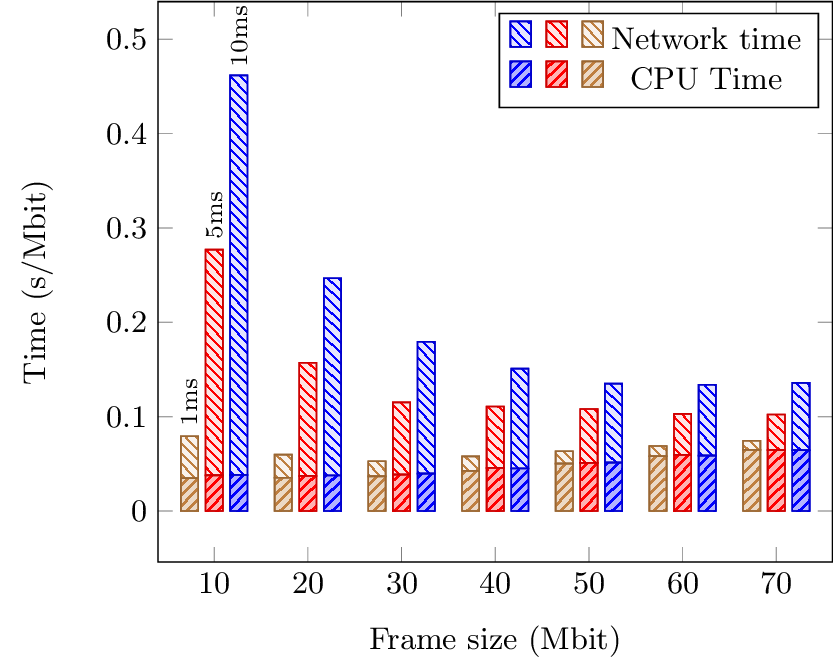}

  \caption{\label{fig:nvscomp}The time spent on correcting a million
    bits of data for different frame sizes. For each frame size, we
    list the time for $1$~ms (first column), $5$~ms (second column),
    and $10$~ms (third column) latency. The contribution of
    communication decreases as the block size increases. For $1$~ms
    latency, we see that the communication becomes an insignificant
    contribution to the overall time spent.}
\end{figure}

It is expected that larger frame sizes will take better advantage of
the parallel binary search for errors, thus reducing the time spent on
communication. However, as can be seen in Fig.~\ref{fig:tpvsn},
after a certain frame size, the throughput drops. This drop is caused
by increased computation time (computation time is super-linear in the
frame size). Fig.~\ref{fig:nvscomp} shows a breakdown of the time
spent on a frame into time spent communicating and time spent
computing. The time spent on communication has an exponential drop-off
which stabilizes after a frame size of 40--50 million bits. The
computation time, on the other hand, increases steadily. From
Fig.~\ref{fig:tpvsn}, we can see that the optimal throughput is
reached for a frame size of 30--40 million bits for the latency of
interest to us ($1$--$3$~ms latency).

The maximum throughput for a single \cascade{} process is 18.97~Mbps
and is reached with a block size of 30~Mbit, a QBER of 1\%, and a
network latency of 1~ms. When the QBER is increased to 5\%, the
throughput is still 12.35~Mbps. Even for 15\% QBER (after which no
secure key can be generated), the throughput is still sufficiently
high for all current high speed QKD systems.

\makeatletter
\newcommand\resetplot{
  \makeatletter
  \pgfplots@stacked@isfirstplottrue
  \makeatother
  \addplot [draw=none] coordinates {(10,0) (20,0) (30,0) (40,0) (50,0) (60,0) (70,0)};
}

The main obstacle in using large frame sizes is memory usage. The
amount of memory used increases linearly in the frame size and
sub-linearly in QBER. Our implementation of \cascade{} uses $303$~Mb
in the optimal scenario of 1\% QBER and 30 million bit frame
size. However, for a QBER of 15\%, the memory usage increases to
$2$~Gb for a frame size of 30 million bits. 

The highest reported throughput achieved with \cascade{} was presented
in \cite{TokyoNet}. The implementation uses $1$ million bit frames and
multiple threads on a quad-core computer (the exact number of threads
used is not reported). It achieves a throughput of $5.5$~Mbps when
correcting for $3.8\%$~QBER using a $45$~km fiber link (at most $1$~ms
latency) for communication. In comparison, our implementation, when
using $1$ million bit frames, achieves $2.6$~Mbps using a single
thread/process when correcting $5\%$~QBER with $1$~ms latency. With 4
\cascade{} processes running in parallel on eight-core computers, we
reach $13.28$~Mbps. As already demonstrated, the optimal frame size is
approximately $30$ million bits. A single process with a $30$ million
bit frame achieves $12.35$~Mbps, while 4 parallel processes using 30
million bit frames reach $27$~Mbps. With 4 processes, however, a
contention for the network occurs between the 4 processes. A frame
size of 15 million bits reduces this contention and results in a
throughput of $36.73$~Mbps.

For comparison with the IR protocol presented in \cite{NIST07}, which
has a throughput of $12.2$~Mbps for a QBER of $1\%$, we ran 4
processes with a frame size of 15 million bits. The throughput when
correcting $1\%$~QBER with $1$~ms latency was $65.99$~Mbps.

By running 8 processes in parallel with a frame size of 10 million
bits, we have achieved a throughput of $82.31$~Mbps when correcting
$1\%$~QBER with a latency of $1$~ms. This is more than four times the
throughput of the previous fastest IR protocol known to
us\cite{ninoqcrypt}.

To demonstrate the performance of our implementation in a realistic,
real-world scenario, we also ran tests on two computers connected by
11~km dedicated fiber, using low cost TP-LINK MC112CS/MC111CS media
converters, converting $100$~Mbps Ethernet connections to a 100Base-FX
fiber connection. The end-to-end latency between Alice and Bob was
approximately $0.4$~ms. Note that most of the latency is in the media
converters, as the signal only takes $0.06$~ms to propagate through
the fiber. To implement an authenticated channel, we created an ssh
tunnel\cite{ssh} between Alice and Bob. This channel is both encrypted
and authenticated with HMAC SHA-2 256. SHA-2 256 is \emph{not} an
information theoretically secure authentication as required for QKD.
Computationally, however, SHA-2 256 is slower than e.g. polynomial
hashing, which is recommended for QKD. The throughput, when correcting
a QBER of $1\%$ using a frame size of 30 million bits, was
$20.74$~Mbps. We repeated the above experiment with 8 parallel
\cascade{} processes, each using a 10 million bit block size. The
throughput of the parallel experiment was $83.49$~Mbps. We note that
by using faster media converters and authentication, the same
throughput could be achieve over much larger distances.

Since we have modified the look-back step of \cascade{}, we confirm
that the efficiency does not deteriorate too much compared to the
standard implementation of \cascade{}. Table~\ref{tab:efficiency}
lists the efficiency of both our implementation and the original
\cascade{} (as reported in \cite{SY00}).  The table clearly shows that
our modification only has a significant influence on efficiency for
high QBER (above 10\%).

\begin{table}
  \caption{\label{tab:efficiency}Efficiency of our implementation of
    \cascade{} (column titled ``this''), the original \cascade{}
    (column tit\-led ``original''), and polar code (column titled
    ``polar'').}
  \newcolumntype{d}{D{.}{.}{1.4}}
  \begin{ruledtabular}
  \begin{tabular}{rddd}
    QBER & \multicolumn{1}{c}{This} & \multicolumn{1}{c}{Original} & \multicolumn{1}{c}{Polar} \\
    \hline
    1\%  & 0.989 & 0.9889 & 0.9875 \\
    3\%  & 0.96 & 0.9602 & 0.975 \\
    5\%  & 0.9231 & 0.9261 & 0.9688 \\
    10\% & 0.7839 & 0.7972 & 0.95 \\
    15\% & 0.5597 & 0.5907 & \multicolumn{1}{c}{--} \\
  \end{tabular}
  \end{ruledtabular}
\end{table}

Table~\ref{tab:efficiency} also lists the efficiency of the polar code
implementation reported in \cite{CNRS13}. The efficiency of the polar
code based IR is higher than that of \cascade{} --- significantly so
after a QBER of 5\%. As pointed out in \cite{CNRS09} and
\cite{WalentaThesis}, LDPC codes can also achieve a higher efficiency
than \cascade{}. When the IR protocol is fast enough to process the
raw key at the rate it arrives, efficiency is more important than
throughput. Does this mean that LDPC or polar code based IR protocols
perform better than \cascade{} when their throughput is high enough?
The answer lies in the remaining term in
Eq.~(\ref{eq:performance}): FER. For \cascade{}, the FER is
negligible. Our implementation of \cascade{} successfully corrected
all $701919$ frames we applied it to in our tests. For the polar code
based IR protocol listed in Table~\ref{tab:efficiency}, the FER is
8\%. For the best LDPC based IR protocol presented in \cite{CNRS13},
the FER is 1\%. When $I_E$ is close to $I(A{:}B)$, efficiency plays a
much more important role than FER. However, when $I_E$ is small
compared to $I(A{:}B)$, FER is the dominating term. This, then,
becomes the scenario where a careful choice between the different IR
protocols must be made.

In our final experiment, we consider the following question: If the
throughput of \cascade{} is significantly higher than what is needed,
can we improve the overall performance by sacrificing surplus
throughput to gain efficiency? One of the contributions in
\cite{Yan08} is to use a different set of partition sizes to improve
the efficiency of \cascade{}. The partition sizes they use are
$0.8/QBER$ for the first round, $4/QBER$ for the second round, and
$n/2$ for another 8 rounds, where $n$ is the frame size. Applying
these partition sizes improves the efficiency, but significantly
decreases throughput and increases FER. To avoid the increased FER and
to increase throughput, we changed the partition size of the third
round to $10^3$, and of the last 7 rounds to $10^6$. Since we use
large frame sizes, the loss in efficiency caused by the extra parity
bits is insignificant. However, the FER becomes negligible. On a
channel with $1$~ms latency and $1\%$~QBER, we achieved $9.47$~Mbps
throughput and an efficiency of $0.9907$ with a frame size of
$30$~million bits. At $5\%$~QBER, the throughput is $6.44$~Mbps and
the efficiency is $0.9465$ using a frame size of $30$ million
bits. This version of \cascade{} has an efficiency close to that of
the LDPC and polar code based IR protocols we have compared to. For
all state-of-the-art QKD systems cited in this paper, the throughput
of this modified \cascade{} is sufficient. If needed, however, by
running two or more processes in parallel, the throughput of this
version of \cascade{} can easily be increased. These last tests
demonstrate that also when throughput is not the limiting factor,
variations of \cascade{} may still be the best performing IR protocol
for many realistic scenarios.

\section{\label{sec:conclusion}Conclusion}

We have demonstrated that, with careful implementation and use,
interactive IR protocols (in particular \cascade{}) can reach
throughput and efficiency sufficiently high for current
state-of-the-art QKD systems. Considering the popularity and
simplicity of \cascade{}, we argue that \cascade{} is a good choice
for QKD implementations in most real-world scenarios. It is, however,
also clear that in settings with extraordinarily high latency on the
classical channel (such as geostationary satellite links), less
interactive IR protocols may be preferable.

Our implementation of \cascade{} has achieved a throughput of
$83.49$~Mbps over a dedicated, authenticated fiber link --- more than
four times faster than has previously been demonstrated by any IR
protocol that we know of. The throughput is an order of magnitude
higher than that needed in state-of-the-art QKD systems. In this case,
the relevant performance metric is Eq.~(\ref{eq:performance}), which
tells us that the overall performance of the system may be improved by
sacrificing surplus throughput to gain efficiency. Even though LDPC
and polar codes have higher efficiency than standard \cascade{} for
large QBER\cite{CNRS09,WalentaThesis}, we have demonstrated that
modified versions of \cascade{}, such as the ones proposed in
\cite{Yan08,SY00}, can reach comparable efficiency while still keeping
throughput high enough.

The main contribution of this paper is to point out that both
communication and computation cost, as well as IR protocol efficiency,
must be carefully considered when choosing an IR protocol for a
specific QKD realization.

Our entire focus has been on the performance of the IR protocol, while
ignoring other design criteria such as the cost of the system. It is
clear that the cost of dedicating a full 8 core CPU, as the one used
in our experiments, to IR is more expensive than an FPGA. We do not
know the performance of \cascade{} when implemented in FPGA. One
challenge with an FPGA implementation of \cascade{} is the large
amount of memory needed. For low-cost systems, a more detailed study
of hardware implementations of \cascade{} is required.

\begin{acknowledgments}
  This work was partially sponsored by the Turkish Republic Ministry
  of Development State Planning Organization --- D.P.T. project
  no.~2009K120200. 
\end{acknowledgments}

\bibliography{bibliography}

\begin{thebibliography}{33}%
\makeatletter
\providecommand \@ifxundefined [1]{%
 \@ifx{#1\undefined}
}%
\providecommand \@ifnum [1]{%
 \ifnum #1\expandafter \@firstoftwo
 \else \expandafter \@secondoftwo
 \fi
}%
\providecommand \@ifx [1]{%
 \ifx #1\expandafter \@firstoftwo
 \else \expandafter \@secondoftwo
 \fi
}%
\providecommand \natexlab [1]{#1}%
\providecommand \enquote  [1]{``#1''}%
\providecommand \bibnamefont  [1]{#1}%
\providecommand \bibfnamefont [1]{#1}%
\providecommand \citenamefont [1]{#1}%
\providecommand \href@noop [0]{\@secondoftwo}%
\providecommand \href [0]{\begingroup \@sanitize@url \@href}%
\providecommand \@href[1]{\@@startlink{#1}\@@href}%
\providecommand \@@href[1]{\endgroup#1\@@endlink}%
\providecommand \@sanitize@url [0]{\catcode `\\12\catcode `\$12\catcode
  `\&12\catcode `\#12\catcode `\^12\catcode `\_12\catcode `\%12\relax}%
\providecommand \@@startlink[1]{}%
\providecommand \@@endlink[0]{}%
\providecommand \url  [0]{\begingroup\@sanitize@url \@url }%
\providecommand \@url [1]{\endgroup\@href {#1}{\urlprefix }}%
\providecommand \urlprefix  [0]{URL }%
\providecommand \Eprint [0]{\href }%
\providecommand \doibase [0]{http://dx.doi.org/}%
\providecommand \selectlanguage [0]{\@gobble}%
\providecommand \bibinfo  [0]{\@secondoftwo}%
\providecommand \bibfield  [0]{\@secondoftwo}%
\providecommand \translation [1]{[#1]}%
\providecommand \BibitemOpen [0]{}%
\providecommand \bibitemStop [0]{}%
\providecommand \bibitemNoStop [0]{.\EOS\space}%
\providecommand \EOS [0]{\spacefactor3000\relax}%
\providecommand \BibitemShut  [1]{\csname bibitem#1\endcsname}%
\let\auto@bib@innerbib\@empty
\bibitem [{\citenamefont {Dixon}\ \emph {et~al.}(2008)\citenamefont {Dixon},
  \citenamefont {Yuan}, \citenamefont {Dynes}, \citenamefont {Sharpe},\ and\
  \citenamefont {Shields}}]{Toshiba08}%
  \BibitemOpen
  \bibfield  {author} {\bibinfo {author} {\bibfnamefont {A.~R.}\ \bibnamefont
  {Dixon}}, \bibinfo {author} {\bibfnamefont {Z.~L.}\ \bibnamefont {Yuan}},
  \bibinfo {author} {\bibfnamefont {J.~F.}\ \bibnamefont {Dynes}}, \bibinfo
  {author} {\bibfnamefont {A.~W.}\ \bibnamefont {Sharpe}}, \ and\ \bibinfo
  {author} {\bibfnamefont {A.~J.}\ \bibnamefont {Shields}},\ }\href@noop {}
  {\bibfield  {journal} {\bibinfo  {journal} {Optics Express}\ }\textbf
  {\bibinfo {volume} {16}},\ \bibinfo {pages} {18790} (\bibinfo {year}
  {2008})}\BibitemShut {NoStop}%
\bibitem [{\citenamefont {Walenta}(2012)}]{ninoqcrypt}%
  \BibitemOpen
  \bibfield  {author} {\bibinfo {author} {\bibfnamefont {N.}~\bibnamefont
  {Walenta}},\ }\href@noop {} {\enquote {\bibinfo {title} {1 {{Mbps}} coherent
  one-way {{QKD}} with dense wavelength division multiplexing and hardware key
  distillation},}\ }\bibinfo {howpublished} {Presentation at QCrypt 2012 ---
  available at http://2012.qcrypt.net/program.html} (\bibinfo {year}
  {2012})\BibitemShut {NoStop}%
\bibitem [{\citenamefont {Martinez-Mateo}\ \emph {et~al.}(2013)\citenamefont
  {Martinez-Mateo}, \citenamefont {Elkouss},\ and\ \citenamefont
  {Martin}}]{MEM13}%
  \BibitemOpen
  \bibfield  {author} {\bibinfo {author} {\bibfnamefont {J.}~\bibnamefont
  {Martinez-Mateo}}, \bibinfo {author} {\bibfnamefont {D.}~\bibnamefont
  {Elkouss}}, \ and\ \bibinfo {author} {\bibfnamefont {V.}~\bibnamefont
  {Martin}},\ }\href@noop {} {\bibfield  {journal} {\bibinfo  {journal}
  {Scientific Reports}\ }\textbf {\bibinfo {volume} {3}},\ \bibinfo {pages}
  {1576} (\bibinfo {year} {2013})}\BibitemShut {NoStop}%
\bibitem [{\citenamefont {Lodewyck}\ \emph {et~al.}(2007)\citenamefont
  {Lodewyck}, \citenamefont {Bloch}, \citenamefont {Garc\'ia-Patr\'on},
  \citenamefont {Fossier}, \citenamefont {Karpov}, \citenamefont {Diamanti},
  \citenamefont {Debuisschert}, \citenamefont {Cerf}, \citenamefont
  {Tualle-Brouri}, \citenamefont {McLaughlin},\ and\ \citenamefont
  {Grangier}}]{CNRS07}%
  \BibitemOpen
  \bibfield  {author} {\bibinfo {author} {\bibfnamefont {J.}~\bibnamefont
  {Lodewyck}}, \bibinfo {author} {\bibfnamefont {M.}~\bibnamefont {Bloch}},
  \bibinfo {author} {\bibfnamefont {R.}~\bibnamefont {Garc\'ia-Patr\'on}},
  \bibinfo {author} {\bibfnamefont {S.}~\bibnamefont {Fossier}}, \bibinfo
  {author} {\bibfnamefont {E.}~\bibnamefont {Karpov}}, \bibinfo {author}
  {\bibfnamefont {E.}~\bibnamefont {Diamanti}}, \bibinfo {author}
  {\bibfnamefont {T.}~\bibnamefont {Debuisschert}}, \bibinfo {author}
  {\bibfnamefont {N.~J.}\ \bibnamefont {Cerf}}, \bibinfo {author}
  {\bibfnamefont {R.}~\bibnamefont {Tualle-Brouri}}, \bibinfo {author}
  {\bibfnamefont {S.~W.}\ \bibnamefont {McLaughlin}}, \ and\ \bibinfo {author}
  {\bibfnamefont {P.}~\bibnamefont {Grangier}},\ }\href {\doibase
  10.1103/PhysRevA.76.042305} {\bibfield  {journal} {\bibinfo  {journal}
  {Physical Review A}\ }\textbf {\bibinfo {volume} {76}},\ \bibinfo {pages}
  {042305} (\bibinfo {year} {2007})}\BibitemShut {NoStop}%
\bibitem [{\citenamefont {Jouguet}\ \emph {et~al.}(2011)\citenamefont
  {Jouguet}, \citenamefont {Kunz-Jacques},\ and\ \citenamefont
  {Leverrier}}]{CNRS11b}%
  \BibitemOpen
  \bibfield  {author} {\bibinfo {author} {\bibfnamefont {P.}~\bibnamefont
  {Jouguet}}, \bibinfo {author} {\bibfnamefont {S.}~\bibnamefont
  {Kunz-Jacques}}, \ and\ \bibinfo {author} {\bibfnamefont {A.}~\bibnamefont
  {Leverrier}},\ }\href {\doibase 10.1103/PhysRevA.84.062317} {\bibfield
  {journal} {\bibinfo  {journal} {Physical Review A}\ }\textbf {\bibinfo
  {volume} {84}},\ \bibinfo {pages} {062317} (\bibinfo {year}
  {2011})}\BibitemShut {NoStop}%
\bibitem [{\citenamefont {Brassard}\ and\ \citenamefont
  {Salvail}(1993)}]{BS93}%
  \BibitemOpen
  \bibfield  {author} {\bibinfo {author} {\bibfnamefont {G.}~\bibnamefont
  {Brassard}}\ and\ \bibinfo {author} {\bibfnamefont {L.}~\bibnamefont
  {Salvail}},\ }in\ \href@noop {} {\emph {\bibinfo {booktitle} {Advances in
  Cryptology — EUROCRYPT '93}}},\ \bibinfo {series and number} {Lecture Notes
  in Computer Science}\ (\bibinfo {year} {1993})\ pp.\ \bibinfo {pages}
  {410--423}\BibitemShut {NoStop}%
\bibitem [{\citenamefont {Jouguet}\ and\ \citenamefont
  {Kunz-Jacques}(2013)}]{CNRS13}%
  \BibitemOpen
  \bibfield  {author} {\bibinfo {author} {\bibfnamefont {P.}~\bibnamefont
  {Jouguet}}\ and\ \bibinfo {author} {\bibfnamefont {S.}~\bibnamefont
  {Kunz-Jacques}},\ }\href@noop {} {\enquote {\bibinfo {title} {High
  performance error correction for quantum key distribution using polar
  codes},}\ } (\bibinfo {year} {2013}),\ \Eprint
  {http://arxiv.org/abs/quant-ph/1204.5882v2} {arXiv:quant-ph/1204.5882v2}
  \BibitemShut {NoStop}%
\bibitem [{\citenamefont {Mink}\ and\ \citenamefont {Nakassis}(2012)}]{NIST12}%
  \BibitemOpen
  \bibfield  {author} {\bibinfo {author} {\bibfnamefont {A.}~\bibnamefont
  {Mink}}\ and\ \bibinfo {author} {\bibfnamefont {A.}~\bibnamefont
  {Nakassis}},\ }\href@noop {} {\bibfield  {journal} {\bibinfo  {journal} {The
  computing science and technology international journal}\ }\textbf {\bibinfo
  {volume} {2}},\ \bibinfo {pages} {6} (\bibinfo {year} {2012})}\BibitemShut
  {NoStop}%
\bibitem [{\citenamefont {Buttler}\ \emph {et~al.}(2003)\citenamefont
  {Buttler}, \citenamefont {Lamoreaux}, \citenamefont {Torgerson},
  \citenamefont {Nickel}, \citenamefont {Donahue},\ and\ \citenamefont
  {Peterson}}]{Winnow}%
  \BibitemOpen
  \bibfield  {author} {\bibinfo {author} {\bibfnamefont {W.~T.}\ \bibnamefont
  {Buttler}}, \bibinfo {author} {\bibfnamefont {S.~K.}\ \bibnamefont
  {Lamoreaux}}, \bibinfo {author} {\bibfnamefont {J.~R.}\ \bibnamefont
  {Torgerson}}, \bibinfo {author} {\bibfnamefont {G.~H.}\ \bibnamefont
  {Nickel}}, \bibinfo {author} {\bibfnamefont {C.~H.}\ \bibnamefont {Donahue}},
  \ and\ \bibinfo {author} {\bibfnamefont {C.~G.}\ \bibnamefont {Peterson}},\
  }\href {\doibase 10.1103/PhysRevA.67.052303} {\bibfield  {journal} {\bibinfo
  {journal} {Physical Review A}\ }\textbf {\bibinfo {volume} {67}},\ \bibinfo
  {pages} {052303} (\bibinfo {year} {2003})}\BibitemShut {NoStop}%
\bibitem [{\citenamefont {Elkouss}\ \emph {et~al.}(2013)\citenamefont
  {Elkouss}, \citenamefont {Martinez-Mateo},\ and\ \citenamefont
  {Martin}}]{MEM13b}%
  \BibitemOpen
  \bibfield  {author} {\bibinfo {author} {\bibfnamefont {D.}~\bibnamefont
  {Elkouss}}, \bibinfo {author} {\bibfnamefont {J.}~\bibnamefont
  {Martinez-Mateo}}, \ and\ \bibinfo {author} {\bibfnamefont {V.}~\bibnamefont
  {Martin}},\ }\href {\doibase 10.1103/PhysRevA.87.042334} {\bibfield
  {journal} {\bibinfo  {journal} {Physical Review A}\ }\textbf {\bibinfo
  {volume} {87}},\ \bibinfo {pages} {042334} (\bibinfo {year}
  {2013})}\BibitemShut {NoStop}%
\bibitem [{\citenamefont {Leverrier}\ \emph {et~al.}(2008)\citenamefont
  {Leverrier}, \citenamefont {All\'eaume}, \citenamefont {Boutros},
  \citenamefont {Z\'emor},\ and\ \citenamefont {Grangier}}]{CNRS08}%
  \BibitemOpen
  \bibfield  {author} {\bibinfo {author} {\bibfnamefont {A.}~\bibnamefont
  {Leverrier}}, \bibinfo {author} {\bibfnamefont {R.}~\bibnamefont
  {All\'eaume}}, \bibinfo {author} {\bibfnamefont {J.}~\bibnamefont {Boutros}},
  \bibinfo {author} {\bibfnamefont {G.}~\bibnamefont {Z\'emor}}, \ and\
  \bibinfo {author} {\bibfnamefont {P.}~\bibnamefont {Grangier}},\ }\href
  {\doibase 10.1103/PhysRevA.77.042325} {\bibfield  {journal} {\bibinfo
  {journal} {Physical Review A}\ }\textbf {\bibinfo {volume} {77}},\ \bibinfo
  {pages} {042325} (\bibinfo {year} {2008})}\BibitemShut {NoStop}%
\bibitem [{\citenamefont {Yan}\ \emph {et~al.}(2008)\citenamefont {Yan},
  \citenamefont {Ren}, \citenamefont {Peng}, \citenamefont {Lin}, \citenamefont
  {Jiang}, \citenamefont {Liu},\ and\ \citenamefont {Guo}}]{Yan08}%
  \BibitemOpen
  \bibfield  {author} {\bibinfo {author} {\bibfnamefont {H.}~\bibnamefont
  {Yan}}, \bibinfo {author} {\bibfnamefont {T.}~\bibnamefont {Ren}}, \bibinfo
  {author} {\bibfnamefont {X.}~\bibnamefont {Peng}}, \bibinfo {author}
  {\bibfnamefont {X.}~\bibnamefont {Lin}}, \bibinfo {author} {\bibfnamefont
  {W.}~\bibnamefont {Jiang}}, \bibinfo {author} {\bibfnamefont
  {T.}~\bibnamefont {Liu}}, \ and\ \bibinfo {author} {\bibfnamefont
  {H.}~\bibnamefont {Guo}},\ }in\ \href@noop {} {\emph {\bibinfo {booktitle}
  {Fourth International Conference on Natural Computation, 2008. ICNC '08}}},\
  Vol.~\bibinfo {volume} {3}\ (\bibinfo {year} {2008})\ pp.\ \bibinfo {pages}
  {637--641}\BibitemShut {NoStop}%
\bibitem [{\citenamefont {Sugimoto}\ and\ \citenamefont
  {Yamazaki}(2000)}]{SY00}%
  \BibitemOpen
  \bibfield  {author} {\bibinfo {author} {\bibfnamefont {T.}~\bibnamefont
  {Sugimoto}}\ and\ \bibinfo {author} {\bibfnamefont {K.}~\bibnamefont
  {Yamazaki}},\ }\href@noop {} {\bibfield  {journal} {\bibinfo  {journal}
  {IEICE Trans. Fundamentals}\ }\textbf {\bibinfo {volume} {E83-A}},\ \bibinfo
  {pages} {1987} (\bibinfo {year} {2000})}\BibitemShut {NoStop}%
\bibitem [{\citenamefont {Liu}\ \emph {et~al.}(2003)\citenamefont {Liu},
  \citenamefont {van Tilborg},\ and\ \citenamefont {van Dijk}}]{LTD03}%
  \BibitemOpen
  \bibfield  {author} {\bibinfo {author} {\bibfnamefont {S.}~\bibnamefont
  {Liu}}, \bibinfo {author} {\bibfnamefont {H.~C.}\ \bibnamefont {van
  Tilborg}}, \ and\ \bibinfo {author} {\bibfnamefont {M.}~\bibnamefont {van
  Dijk}},\ }\href@noop {} {\bibfield  {journal} {\bibinfo  {journal} {Designs,
  Codes and Cryptography}\ }\textbf {\bibinfo {volume} {30}},\ \bibinfo {pages}
  {39} (\bibinfo {year} {2003})}\BibitemShut {NoStop}%
\bibitem [{\citenamefont {Mink}(2007)}]{NIST07}%
  \BibitemOpen
  \bibfield  {author} {\bibinfo {author} {\bibfnamefont {A.}~\bibnamefont
  {Mink}},\ }in\ \href@noop {} {\emph {\bibinfo {booktitle} {Proc. SPIE 6780,
  Quantum Communications Realized}}}\ (\bibinfo {year} {2007})\ p.\ \bibinfo
  {pages} {678014}\BibitemShut {NoStop}%
\bibitem [{\citenamefont {Patel}\ \emph {et~al.}(2012)\citenamefont {Patel},
  \citenamefont {Dynes}, \citenamefont {Choi}, \citenamefont {Sharpe},
  \citenamefont {Dixon}, \citenamefont {Yuan}, \citenamefont {Penty}, ,\ and\
  \citenamefont {Shields}}]{Toshiba12}%
  \BibitemOpen
  \bibfield  {author} {\bibinfo {author} {\bibfnamefont {K.~A.}\ \bibnamefont
  {Patel}}, \bibinfo {author} {\bibfnamefont {J.~F.}\ \bibnamefont {Dynes}},
  \bibinfo {author} {\bibfnamefont {I.}~\bibnamefont {Choi}}, \bibinfo {author}
  {\bibfnamefont {A.~W.}\ \bibnamefont {Sharpe}}, \bibinfo {author}
  {\bibfnamefont {A.~R.}\ \bibnamefont {Dixon}}, \bibinfo {author}
  {\bibfnamefont {Z.~L.}\ \bibnamefont {Yuan}}, \bibinfo {author}
  {\bibfnamefont {R.~V.}\ \bibnamefont {Penty}}, , \ and\ \bibinfo {author}
  {\bibfnamefont {A.~J.}\ \bibnamefont {Shields}},\ }\href {\doibase
  10.1103/PhysRevX.2.041010} {\bibfield  {journal} {\bibinfo  {journal}
  {Physical Review X}\ }\textbf {\bibinfo {volume} {2}},\ \bibinfo {pages}
  {041010} (\bibinfo {year} {2012})}\BibitemShut {NoStop}%
\bibitem [{\citenamefont {Nauerth}\ \emph {et~al.}(2013)\citenamefont
  {Nauerth}, \citenamefont {Moll}, \citenamefont {Rau}, \citenamefont {Fuchs},
  \citenamefont {Horwath}, \citenamefont {Frick},\ and\ \citenamefont
  {Weinfurter}}]{Fuchs+13}%
  \BibitemOpen
  \bibfield  {author} {\bibinfo {author} {\bibfnamefont {S.}~\bibnamefont
  {Nauerth}}, \bibinfo {author} {\bibfnamefont {F.}~\bibnamefont {Moll}},
  \bibinfo {author} {\bibfnamefont {M.}~\bibnamefont {Rau}}, \bibinfo {author}
  {\bibfnamefont {C.}~\bibnamefont {Fuchs}}, \bibinfo {author} {\bibfnamefont
  {J.}~\bibnamefont {Horwath}}, \bibinfo {author} {\bibfnamefont
  {S.}~\bibnamefont {Frick}}, \ and\ \bibinfo {author} {\bibfnamefont
  {H.}~\bibnamefont {Weinfurter}},\ }\href@noop {} {\bibfield  {journal}
  {\bibinfo  {journal} {Nature Photonics}\ }\textbf {\bibinfo {volume} {7}},\
  \bibinfo {pages} {382–} (\bibinfo {year} {2013})}\BibitemShut {NoStop}%
\bibitem [{\citenamefont {Walenta}(2013)}]{WalentaThesis}%
  \BibitemOpen
  \bibfield  {author} {\bibinfo {author} {\bibfnamefont {N.}~\bibnamefont
  {Walenta}},\ }\emph {\bibinfo {title} {Concepts, components and
  implementations for quantum key distribution over optical fiber}},\
  \href@noop {} {Ph.D. thesis},\ \bibinfo  {school} {Universit\'e de Gen\`eve}
  (\bibinfo {year} {2013})\BibitemShut {NoStop}%
\bibitem [{\citenamefont {Tang}\ \emph {et~al.}(2006)\citenamefont {Tang},
  \citenamefont {Ma}, \citenamefont {Mink}, \citenamefont {Nakassis},
  \citenamefont {Xu}, \citenamefont {Hershman}, \citenamefont {Bienfang},
  \citenamefont {Su}, \citenamefont {Boisvert}, \citenamefont {Clark},\ and\
  \citenamefont {Williams}}]{NIST06}%
  \BibitemOpen
  \bibfield  {author} {\bibinfo {author} {\bibfnamefont {X.}~\bibnamefont
  {Tang}}, \bibinfo {author} {\bibfnamefont {L.}~\bibnamefont {Ma}}, \bibinfo
  {author} {\bibfnamefont {A.}~\bibnamefont {Mink}}, \bibinfo {author}
  {\bibfnamefont {A.}~\bibnamefont {Nakassis}}, \bibinfo {author}
  {\bibfnamefont {H.}~\bibnamefont {Xu}}, \bibinfo {author} {\bibfnamefont
  {B.}~\bibnamefont {Hershman}}, \bibinfo {author} {\bibfnamefont
  {J.}~\bibnamefont {Bienfang}}, \bibinfo {author} {\bibfnamefont
  {D.}~\bibnamefont {Su}}, \bibinfo {author} {\bibfnamefont {R.~F.}\
  \bibnamefont {Boisvert}}, \bibinfo {author} {\bibfnamefont {C.}~\bibnamefont
  {Clark}}, \ and\ \bibinfo {author} {\bibfnamefont {C.}~\bibnamefont
  {Williams}},\ }in\ \href@noop {} {\emph {\bibinfo {booktitle} {Proc. SPIE
  6244, Quantum Information and Computation IV}}}\ (\bibinfo {year} {2006})\
  p.\ \bibinfo {pages} {62440P}\BibitemShut {NoStop}%
\bibitem [{\citenamefont {Dixon}\ \emph {et~al.}(2010)\citenamefont {Dixon},
  \citenamefont {Yuan}, \citenamefont {Dynes}, \citenamefont {Sharpe},\ and\
  \citenamefont {Shields}}]{Toshiba10}%
  \BibitemOpen
  \bibfield  {author} {\bibinfo {author} {\bibfnamefont {A.~R.}\ \bibnamefont
  {Dixon}}, \bibinfo {author} {\bibfnamefont {Z.~L.}\ \bibnamefont {Yuan}},
  \bibinfo {author} {\bibfnamefont {J.~F.}\ \bibnamefont {Dynes}}, \bibinfo
  {author} {\bibfnamefont {A.~W.}\ \bibnamefont {Sharpe}}, \ and\ \bibinfo
  {author} {\bibfnamefont {A.~J.}\ \bibnamefont {Shields}},\ }\href@noop {}
  {\bibfield  {journal} {\bibinfo  {journal} {Applied Physics Letter}\ }\textbf
  {\bibinfo {volume} {96}},\ \bibinfo {pages} {161102} (\bibinfo {year}
  {2010})}\BibitemShut {NoStop}%
\bibitem [{\citenamefont {Bennett}\ and\ \citenamefont
  {Brassard}(1984)}]{BB84}%
  \BibitemOpen
  \bibfield  {author} {\bibinfo {author} {\bibfnamefont {C.~H.}\ \bibnamefont
  {Bennett}}\ and\ \bibinfo {author} {\bibfnamefont {G.}~\bibnamefont
  {Brassard}},\ }in\ \href@noop {} {\emph {\bibinfo {booktitle} {Proceedings of
  the IEEE International Conference on Computers, Systems and Signal
  Processing}}}\ (\bibinfo {year} {1984})\ pp.\ \bibinfo {pages}
  {175--179}\BibitemShut {NoStop}%
\bibitem [{\citenamefont {Hwang}(2003)}]{H03}%
  \BibitemOpen
  \bibfield  {author} {\bibinfo {author} {\bibfnamefont {W.-Y.}\ \bibnamefont
  {Hwang}},\ }\href@noop {} {\bibfield  {journal} {\bibinfo  {journal}
  {Physical Review Letters}\ }\textbf {\bibinfo {volume} {91}},\ \bibinfo
  {pages} {057901} (\bibinfo {year} {2003})}\BibitemShut {NoStop}%
\bibitem [{\citenamefont {Yuan}\ \emph {et~al.}(2007)\citenamefont {Yuan},
  \citenamefont {Sharpe},\ and\ \citenamefont {Shields}}]{Toshiba07}%
  \BibitemOpen
  \bibfield  {author} {\bibinfo {author} {\bibfnamefont {Z.~L.}\ \bibnamefont
  {Yuan}}, \bibinfo {author} {\bibfnamefont {A.~W.}\ \bibnamefont {Sharpe}}, \
  and\ \bibinfo {author} {\bibfnamefont {A.~J.}\ \bibnamefont {Shields}},\
  }\href@noop {} {\bibfield  {journal} {\bibinfo  {journal} {Applied Physics
  Letters}\ }\textbf {\bibinfo {volume} {90}},\ \bibinfo {pages} {011118}
  (\bibinfo {year} {2007})}\BibitemShut {NoStop}%
\bibitem [{\citenamefont {Stucki}\ \emph {et~al.}(2005)\citenamefont {Stucki},
  \citenamefont {Brunner}, \citenamefont {Gisin}, \citenamefont {Scarani},\
  and\ \citenamefont {Zbinden}}]{SBGSZ05}%
  \BibitemOpen
  \bibfield  {author} {\bibinfo {author} {\bibfnamefont {D.}~\bibnamefont
  {Stucki}}, \bibinfo {author} {\bibfnamefont {N.}~\bibnamefont {Brunner}},
  \bibinfo {author} {\bibfnamefont {N.}~\bibnamefont {Gisin}}, \bibinfo
  {author} {\bibfnamefont {V.}~\bibnamefont {Scarani}}, \ and\ \bibinfo
  {author} {\bibfnamefont {H.}~\bibnamefont {Zbinden}},\ }\href@noop {}
  {\bibfield  {journal} {\bibinfo  {journal} {Applied Physical Letter}\
  }\textbf {\bibinfo {volume} {87}},\ \bibinfo {pages} {194108} (\bibinfo
  {year} {2005})}\BibitemShut {NoStop}%
\bibitem [{\citenamefont {Walenta}\ \emph {et~al.}(2012)\citenamefont
  {Walenta}, \citenamefont {Lunghi}, \citenamefont {Guinnard}, \citenamefont
  {Houlmann}, \citenamefont {Zbinden},\ and\ \citenamefont {Gisin}}]{Geneva12}%
  \BibitemOpen
  \bibfield  {author} {\bibinfo {author} {\bibfnamefont {N.}~\bibnamefont
  {Walenta}}, \bibinfo {author} {\bibfnamefont {T.}~\bibnamefont {Lunghi}},
  \bibinfo {author} {\bibfnamefont {O.}~\bibnamefont {Guinnard}}, \bibinfo
  {author} {\bibfnamefont {R.}~\bibnamefont {Houlmann}}, \bibinfo {author}
  {\bibfnamefont {H.}~\bibnamefont {Zbinden}}, \ and\ \bibinfo {author}
  {\bibfnamefont {N.}~\bibnamefont {Gisin}},\ }\href@noop {} {\bibfield
  {journal} {\bibinfo  {journal} {Journal of Applied Physics}\ }\textbf
  {\bibinfo {volume} {116}},\ \bibinfo {pages} {063106} (\bibinfo {year}
  {2012})}\BibitemShut {NoStop}%
\bibitem [{\citenamefont {Wang}\ \emph {et~al.}(2012)\citenamefont {Wang},
  \citenamefont {Chen}, \citenamefont {Guo}, \citenamefont {Yin}, \citenamefont
  {Li}, \citenamefont {Zhou}, \citenamefont {Guo},\ and\ \citenamefont
  {Han}}]{wcg12}%
  \BibitemOpen
  \bibfield  {author} {\bibinfo {author} {\bibfnamefont {S.}~\bibnamefont
  {Wang}}, \bibinfo {author} {\bibfnamefont {W.}~\bibnamefont {Chen}}, \bibinfo
  {author} {\bibfnamefont {J.-F.}\ \bibnamefont {Guo}}, \bibinfo {author}
  {\bibfnamefont {Z.-Q.}\ \bibnamefont {Yin}}, \bibinfo {author} {\bibfnamefont
  {H.-W.}\ \bibnamefont {Li}}, \bibinfo {author} {\bibfnamefont
  {Z.}~\bibnamefont {Zhou}}, \bibinfo {author} {\bibfnamefont {G.-C.}\
  \bibnamefont {Guo}}, \ and\ \bibinfo {author} {\bibfnamefont {Z.-F.}\
  \bibnamefont {Han}},\ }\href@noop {} {\bibfield  {journal} {\bibinfo
  {journal} {Optics Letters}\ }\textbf {\bibinfo {volume} {37}},\ \bibinfo
  {pages} {1008} (\bibinfo {year} {2012})}\BibitemShut {NoStop}%
\bibitem [{\citenamefont {Sasaki}\ \emph {et~al.}(2011)\citenamefont {Sasaki},
  \citenamefont {Fujiwara}, \citenamefont {Ishizuka}, \citenamefont {Klaus},
  \citenamefont {Wakui}, \citenamefont {Takeoka}, \citenamefont {Miki},
  \citenamefont {Yamashita}, \citenamefont {Wang}, \citenamefont {Tanaka},
  \citenamefont {Yoshino}, \citenamefont {Nambu}, \citenamefont {Takahashi},
  \citenamefont {Tajima}, \citenamefont {Tomita}, \citenamefont {Domeki},
  \citenamefont {Hasegawa}, \citenamefont {Sakai}, \citenamefont {Kobayashi},
  \citenamefont {Asai}, \citenamefont {Shimizu}, \citenamefont {Tokura},
  \citenamefont {Tsurumaru}, \citenamefont {Matsui}, \citenamefont {Honjo},
  \citenamefont {Tamaki}, \citenamefont {Takesue}, \citenamefont {Tokura},
  \citenamefont {Dynes}, \citenamefont {Dixon}, \citenamefont {Sharpe},
  \citenamefont {Yuan}, \citenamefont {Shields}, \citenamefont {Uchikoga},
  \citenamefont {Legré}, \citenamefont {Robyr}, \citenamefont {Trinkler},
  \citenamefont {Monat}, \citenamefont {Page}, \citenamefont {Ribordy},
  \citenamefont {Poppe}, \citenamefont {Allacher}, \citenamefont {Maurhart},
  \citenamefont {L\"anger}, \citenamefont {Peev},\ and\ \citenamefont
  {Zeilinger}}]{TokyoNet}%
  \BibitemOpen
  \bibfield  {author} {\bibinfo {author} {\bibfnamefont {M.}~\bibnamefont
  {Sasaki}}, \bibinfo {author} {\bibfnamefont {M.}~\bibnamefont {Fujiwara}},
  \bibinfo {author} {\bibfnamefont {H.}~\bibnamefont {Ishizuka}}, \bibinfo
  {author} {\bibfnamefont {W.}~\bibnamefont {Klaus}}, \bibinfo {author}
  {\bibfnamefont {K.}~\bibnamefont {Wakui}}, \bibinfo {author} {\bibfnamefont
  {M.}~\bibnamefont {Takeoka}}, \bibinfo {author} {\bibfnamefont
  {S.}~\bibnamefont {Miki}}, \bibinfo {author} {\bibfnamefont {T.}~\bibnamefont
  {Yamashita}}, \bibinfo {author} {\bibfnamefont {Z.}~\bibnamefont {Wang}},
  \bibinfo {author} {\bibfnamefont {A.}~\bibnamefont {Tanaka}}, \bibinfo
  {author} {\bibfnamefont {K.}~\bibnamefont {Yoshino}}, \bibinfo {author}
  {\bibfnamefont {Y.}~\bibnamefont {Nambu}}, \bibinfo {author} {\bibfnamefont
  {S.}~\bibnamefont {Takahashi}}, \bibinfo {author} {\bibfnamefont
  {A.}~\bibnamefont {Tajima}}, \bibinfo {author} {\bibfnamefont
  {A.}~\bibnamefont {Tomita}}, \bibinfo {author} {\bibfnamefont
  {T.}~\bibnamefont {Domeki}}, \bibinfo {author} {\bibfnamefont
  {T.}~\bibnamefont {Hasegawa}}, \bibinfo {author} {\bibfnamefont
  {Y.}~\bibnamefont {Sakai}}, \bibinfo {author} {\bibfnamefont
  {H.}~\bibnamefont {Kobayashi}}, \bibinfo {author} {\bibfnamefont
  {T.}~\bibnamefont {Asai}}, \bibinfo {author} {\bibfnamefont {K.}~\bibnamefont
  {Shimizu}}, \bibinfo {author} {\bibfnamefont {T.}~\bibnamefont {Tokura}},
  \bibinfo {author} {\bibfnamefont {T.}~\bibnamefont {Tsurumaru}}, \bibinfo
  {author} {\bibfnamefont {M.}~\bibnamefont {Matsui}}, \bibinfo {author}
  {\bibfnamefont {T.}~\bibnamefont {Honjo}}, \bibinfo {author} {\bibfnamefont
  {K.}~\bibnamefont {Tamaki}}, \bibinfo {author} {\bibfnamefont
  {H.}~\bibnamefont {Takesue}}, \bibinfo {author} {\bibfnamefont
  {Y.}~\bibnamefont {Tokura}}, \bibinfo {author} {\bibfnamefont {J.~F.}\
  \bibnamefont {Dynes}}, \bibinfo {author} {\bibfnamefont {A.~R.}\ \bibnamefont
  {Dixon}}, \bibinfo {author} {\bibfnamefont {A.~W.}\ \bibnamefont {Sharpe}},
  \bibinfo {author} {\bibfnamefont {Z.~L.}\ \bibnamefont {Yuan}}, \bibinfo
  {author} {\bibfnamefont {A.~J.}\ \bibnamefont {Shields}}, \bibinfo {author}
  {\bibfnamefont {S.}~\bibnamefont {Uchikoga}}, \bibinfo {author}
  {\bibfnamefont {M.}~\bibnamefont {Legré}}, \bibinfo {author} {\bibfnamefont
  {S.}~\bibnamefont {Robyr}}, \bibinfo {author} {\bibfnamefont
  {P.}~\bibnamefont {Trinkler}}, \bibinfo {author} {\bibfnamefont
  {L.}~\bibnamefont {Monat}}, \bibinfo {author} {\bibfnamefont {J.-B.}\
  \bibnamefont {Page}}, \bibinfo {author} {\bibfnamefont {G.}~\bibnamefont
  {Ribordy}}, \bibinfo {author} {\bibfnamefont {A.}~\bibnamefont {Poppe}},
  \bibinfo {author} {\bibfnamefont {A.}~\bibnamefont {Allacher}}, \bibinfo
  {author} {\bibfnamefont {O.}~\bibnamefont {Maurhart}}, \bibinfo {author}
  {\bibfnamefont {T.}~\bibnamefont {L\"anger}}, \bibinfo {author}
  {\bibfnamefont {M.}~\bibnamefont {Peev}}, \ and\ \bibinfo {author}
  {\bibfnamefont {A.}~\bibnamefont {Zeilinger}},\ }\href@noop {} {\bibfield
  {journal} {\bibinfo  {journal} {Optics Express}\ }\textbf {\bibinfo {volume}
  {19}},\ \bibinfo {pages} {10387} (\bibinfo {year} {2011})}\BibitemShut
  {NoStop}%
\bibitem [{Note1()}]{Note1}%
  \BibitemOpen
  \bibinfo {note} {Citation to \cite {NIST07} in \cite {NIST12} leads us to
  believe that the IR protocol used in \cite {NIST07} is \protect \textsc
  {cascade}{}, or a variant thereof, which would make it the fastest
  implementation of \protect \textsc {cascade}{} known to us.}\BibitemShut
  {Stop}%
\bibitem [{\citenamefont {Peev}(2013)}]{louis}%
  \BibitemOpen
  \bibfield  {author} {\bibinfo {author} {\bibfnamefont {M.}~\bibnamefont
  {Peev}},\ }\href@noop {} {}\bibinfo {howpublished} {private communication}
  (\bibinfo {year} {2013})\BibitemShut {NoStop}%
\bibitem [{\citenamefont {{Austrian Institute of Technology}}(2013)}]{aitqkd}%
  \BibitemOpen
  \bibfield  {author} {\bibinfo {author} {\bibnamefont {{Austrian Institute of
  Technology}}},\ }\href@noop {} {\enquote {\bibinfo {title} {{AIT QKD
  Software}},}\ }\bibinfo {howpublished}
  {http://sqt.ait.ac.at/software/qkd-software} (\bibinfo {year} {accessed July
  15, 2013})\BibitemShut {NoStop}%
\bibitem [{\citenamefont {{The Linux Foundation}}(2013)}]{linuxtc}%
  \BibitemOpen
  \bibfield  {author} {\bibinfo {author} {\bibnamefont {{The Linux
  Foundation}}},\ }\href@noop {} {\enquote {\bibinfo {title} {Netem},}\
  }\bibinfo {howpublished}
  {http://www.linuxfoundation.org/collaborate/work\-groups/\-networking/netem}
  (\bibinfo {year} {accessed July 9, 2013})\BibitemShut {NoStop}%
\bibitem [{\citenamefont {{OpenSSH}}(2013)}]{ssh}%
  \BibitemOpen
  \bibfield  {author} {\bibinfo {author} {\bibnamefont {{OpenSSH}}},\
  }\href@noop {} {}\bibinfo {howpublished} {http://www.openssh.org} (\bibinfo
  {year} {accessed July 15, 2013})\BibitemShut {NoStop}%
\bibitem [{\citenamefont {Elkouss}\ \emph {et~al.}(2009)\citenamefont
  {Elkouss}, \citenamefont {Leverrier}, \citenamefont {All\'eaume},\ and\
  \citenamefont {Boutros}}]{CNRS09}%
  \BibitemOpen
  \bibfield  {author} {\bibinfo {author} {\bibfnamefont {D.}~\bibnamefont
  {Elkouss}}, \bibinfo {author} {\bibfnamefont {A.}~\bibnamefont {Leverrier}},
  \bibinfo {author} {\bibfnamefont {R.}~\bibnamefont {All\'eaume}}, \ and\
  \bibinfo {author} {\bibfnamefont {J.~J.}\ \bibnamefont {Boutros}},\ }in\
  \href@noop {} {\emph {\bibinfo {booktitle} {IEEE International Symposium on
  Information Theory, 2009. ISIT 2009}}}\ (\bibinfo {year} {2009})\ pp.\
  \bibinfo {pages} {1879--1883}\BibitemShut {NoStop}%
\end{thebibliography}%

\end{document}